# *HST* Observations of the Distant Cluster 0016+16: Quantitative Morphology of Confirmed Cluster Members[1,2]

Gregory D. Wirth and David C. Koo
University of California Observatories/Lick Observatory,
Board of Studies in Astronomy and Astrophysics
University of California, Santa Cruz, California 95064

and

Richard G. Kron
Fermi National Accelerator Laboratory
MS 127, Box 500, Batavia, Illinois 60510

## ABSTRACT

We present *HST* images of 24 confirmed members of the distant galaxy cluster Cl0016+16 at redshift 0.55. The Balmer-strong ("E+A") and emission-line galaxies frequently show unusual visual morphology, implying that galaxian interactions produce "active" galaxies in moderate-$z$ clusters. We use the image concentration index as a quantitative measure of morphology to show that these unusual galaxies appear disklike, while the normal red galaxies resemble E/S0s. Although consistent with *HST* observations by Dressler *et al.* in Cl0939+4713 ($z = 0.41$), our results differ from the Couch *et al.* finding that most Balmer-strong galaxies in AC114 ($z = 0.31$) resemble ellipticals. The entire "E+A" sample is small, but if future studies confirm their diversity, it will suggest that they have different origins.

*Subject headings:*   galaxies: clusters of — galaxies: evolution — galaxies: photometry — galaxies: structure

## 1. Introduction

The galaxy cluster 0016+16 is noteworthy for lying at a relatively high redshift (0.55) and being very rich (Koo 1981). It is among the three most luminous x-ray clusters known (Henry *et al.* 1992) and is one of the few clusters exhibiting a microwave decrement due to the Sunyaev-Zeldovich effect (Lasenby 1992). The galaxies in Cl0016+16 are enigmatic: photometrically, they are virtually all red (Koo 1981, Ellis *et al.* 1985, Aragón-Salamanca *et al.* 1993) and thus provide a counter-example of the "Butcher-Oemler effect" (Butcher & Oemler 1984); yet spectroscopic evidence of recent star formation found in many of the red galaxies puts the fraction of "active" galaxies in Cl0016+16 among the highest known (Dressler & Gunn 1992; hereafter DG). To complement these extensive observations of Cl0016+16, we obtained *Hubble Space Telescope (HST)* data to investigate galaxy morphology. This *Letter* presents the first results from these high-resolution images of Cl0016+16, and compares them to findings in lower-redshift clusters.

## 2. Observations

The *HST* Wide-Field Camera (WFC-1) observed Cl0016+16 on 1993 January 3 for a total of 10,500 s through the $F785LP$ filter ($\lambda_{\rm eff} \approx 8900$Å) as part of program GTO 3685. A mean image was computed from 5 separate exposures using a cosmic-ray rejection filter with $2\sigma$ threshold to eliminate cosmic ray events. Applying a flat-field correction image (Phillips *et al.* 1994) minimized residual artificial structure not removed in the WFC pipeline reduction. CCD columns which were systematically too high or low by only a few ADU were corrected by addition of a constant value. Highly deviant columns were eliminated by interpolation, as were pixels with values more than $7.5\sigma$ below sky. We located and removed "hot pixels" on the CCD with a cosmic-ray removal algorithm, while verifying that object cores were not mistakenly replaced.

We corrected for the spherical aberration present in the WFC images using 16 iterations of the Lucy-Richardson image restoration algorithm and a point-spread function (PSF) typical of a K star (generated using TinyTim version 2.2; Krist 1992). Deconvolution significantly reduced the power in the wings of the PSF: for 4 confirmed stars in our frames, the enclosed light within $0\rlap.{''}5$ increased from 37% to 85%. Figure 1 (Plate ??) shows the deconvolved images for confirmed members of Cl0016+16 lying within the WFC field.

Figure 1 available from author via e-mail to wirth@lick.ucsc.edu

Fig. 1.— A montage showing deconvolved images of the 24 confirmed members of Cl0016+16 appearing in the WFC images. The ordering runs to the right and down, with objects grouped first by spectral class, and within classes by Petrosian magnitude, $m_P$. Each galaxy is labeled with its catalog number and spectral classification (DG). North is at 1 o'clock, and east is at 10 o'clock for each of the images, which all are 25 pixels ($2\farcs5$, 21 kpc for $z = 0.55, q_0 = 0.05, h = 0.5$) per side. Contours beginning at $3\sigma$ above sky (surface brightness $\mu(\text{F786LP}) \approx 22.0$ mag arcsec$^{-2}$) and spaced at intervals of 0.5 mag in surface brightness are superimposed on greyscale images stretching logarithmically from $1\sigma$ to $10\sigma$ above sky. Note that interpolation was used to remove a bad column running (horizontally) through object #181, and that #224 lies near the edge of the illuminated region of the CCD, resulting in the artificial cutoff at the right-hand side.

## 3. Analysis

The deconvolved WFC frames contain images of 170 objects to a limiting magnitude of $m_{\rm lim} \approx 24$ in the $F785LP$ passband, corresponding to $L \approx L^* + 2$ in Cl0016+16 for $q_0 = 0.05$. Although we expect most objects to belong to the $z = 0.55$ cluster, only 30 are redshift-confirmed members (DG), and 24 of them lie within our field of view. We now consider the properties of these secure cluster members.

### 3.1. Qualitative Morphology

Comparing the morphologies of Cl0016+16's galaxies to those in lower-redshift clusters (Couch *et al.* 1994, Dressler *et al.* 1994) may yield insight into galaxy evolution, the Butcher-Oemler effect, and the unusual "E+A" population. Unfortunately, the low signal-to-noise ratio (SNR) in our images precluded reliable visual classifications; thus, for morphology we relied on quantitative measures of galaxy shape. Still, we sought visual evidence for galaxy interactions following the prescription of Couch *et al.*

In their study of intermediate-redshift clusters with *HST*, Couch *et al.* identified recent mergers or interactions by the presence of: (1) a distinct double nucleus within a galaxy; (2) comparable-brightness neighbors connected by tidal tails; and (3) satellite galaxies. Although our WFC images have poorer physical resolution and SNR than the Couch *et al.* images, we applied similar visual criteria by identifying double nuclei, tidal tails, companion objects fainter than the primary within 2″, and image substructure. Elongated objects were also noted, regardless of their association with mergers or interactions. This qualitative information on the cluster galaxies appears in the last column of Table 1.

### 3.2. Quantitative Morphology

To augment our visual evaluation of the Cl0016+16 galaxies, we measured their size, surface brightness, and profile shape. First, we derived growth curves within circular apertures using algorithms generously shared by M. A. Bershady. By ignoring neighboring objects when obtaining a galaxy's profile, the software yields reasonably accurate photometry in crowded fields such as this rich cluster core.

We quantified profile shapes using the image concentration index, $C$. To measure $C$ each galaxy must have a well-defined "total light," difficult to measure in our crowded,

TABLE 1

Properties of Cl0016+16 Members

| Galaxy (1) | $\alpha$ (1950) (2) | $\delta$ (1950) (3) | Chip (4) | $x_c, y_c$ (5) | $C$ (6) | $m_P$ (7) | $g-r$ (8) | Spectrum (9) | Image (10) |
|---|---|---|---|---|---|---|---|---|---|
| 106 | $0^h 16^m 00^s.99$ | $+16°08'26''.9$ | 2 | 668, 665 | 3.1 | 21.5 | 1.4 | a | N |
| 114 | 0 16 00.63 | +16 08 29.6 | 2 | 613, 660 | 3.0 | 20.2 | 1.6 | k | E |
| 134 | 0 16 01.61 | +16 08 37.6 | 2 | 713, 537 | 2.8 | 21.3 | 0.9 | A | C |
| 139 | 0 15 57.25 | +16 08 39.6 | 2 | 129, 743 | 2.8 | 20.7 | 1.2 | a | N |
| 159 | 0 15 54.73 | +16 08 50.0 | 3 | 774, 281 | 4.2 | 20.8 | 1.4 | k | N |
| 163 | 0 16 01.00 | +16 08 52.1 | 2 | 579, 435 | 2.6 | 21.4 | 1.4 | a | S: |
| 181 | 0 15 55.75 | +16 09 01.6 | 3 | 614, 189 | $3.3^a$ | $20.4^a$ | 1.5 | A | $N^a$ |
| 186 | 0 15 56.03 | +16 09 04.3 | 3 | 575, 164 | 3.0 | 19.3 | 1.2 | k,e | E,C |
| 205a | 0 15 57.98 | +16 09 13.4 | 2 | 107, 401 | 3.6 | 19.5 | 1.8 | k | N |
| 224 | 0 15 57.64 | +16 09 23.8 | 2 | 25, 318 | 3.4 | 18.7 | 1.7 | k | N |
| 251 | 0 15 58.40 | +16 09 36.0 | 2 | 81, 167 | 3.6 | 18.1 | 1.7 | k | C |
| 256 | 0 15 58.72 | +16 09 39.3 | 2 | 111, 120 | 3.5 | 18.0 | 1.8 | k | C: |
| 264 | 0 15 55.93 | +16 09 41.6 | 3 | 241, 309 | 4.4 | 19.5 | 1.8 | k | N |
| 269 | 0 15 59.02 | +16 09 42.6 | 2 | 140, 74 | 4.5 | 19.0 | 1.6 | k | C |
| 300 | 0 16 00.20 | +16 09 55.5 | 1 | 158, 264 | 4.0 | 20.8 | 1.3 | k | N |
| 304 | 0 15 56.32 | +16 09 58.9 | 3 | 62, 319 | 4.5 | 19.5 | 1.6 | k | N |
| 320 | 0 15 55.23 | +16 10 05.0 | 3 | 61, 487 | 4.4 | 20.4 | 1.8 | k | N |
| 338 | 0 16 00.81 | +16 10 12.0 | 1 | 341, 289 | 2.2 | 20.9 | 1.1 | a,e | D |
| 352 | 0 15 56.38 | +16 10 16.5 | 4 | 368, 160 | 3.8 | 20.6 | 1.5 | a: | S: |
| 356 | 0 15 54.30 | +16 10 17.8 | 4 | 645, 62 | 3.5 | 21.6 | 1.0 | e | E |
| 371 | 0 16 03.74 | +16 10 23.6 | 1 | 594, 635 | 2.5 | 19.4 | 1.2 | A | D,S |
| 411 | 0 15 57.77 | +16 10 42.2 | 4 | 277, 464 | 2.8 | 20.7 | 1.4 | A | C,T |
| 413 | 0 15 59.99 | +16 10 42.9 | 1 | 580, 71 | 3.3 | 21.1 | 0.6 | e | T: |
| 439 | 0 15 57.85 | +16 11 08.3 | 4 | 361, 707 | 3.3 | 19.6 | 1.7 | k | N |

[a] Object occupies bad columns in the WFC image and thus values are suspect.

Table 1: Col. (1) Object designation from DG; cols. (2)–(3) Object center, computed from DG offsets; col. (4) CCD on which object lies; col. (5) Location of galaxy center on WFC image; col. (6) Concentration index measured from the deconvolved image; col. (7) Petrosian magnitude in F785LP passband; col. (8) Ground-based color from DG; col. (9) Spectroscopic classification from DG, where "a" denotes "E+A"-type spectrum, "A" indicates "E+A"-type spectrum with strong Balmer lines, "k" signifies K-star spectrum typical of early-type galaxies, "e" connotes emission-line spectrum, and colon implies uncertainty; col. (10) Keywords characterizing the WFC images: "C" connotes companions within 2″, "D" denotes a double nucleus, "E" identifies an elongated image, "S" stands for substructure, "T" specifies the presence of spiral arms or tidal tails, and "N" indicates no unusual morphology.

low-SNR data. We instead defined "Petrosian magnitudes" within a radius for which the function $\eta \equiv I(r)/\langle I(r)\rangle$ equals 0.1 (Petrosian 1976). $I(r)$ is the surface brightness at radius $r$, and $\langle I(r)\rangle$ is the *average* surface brightness (flux units) within the same radius. Such "metric" sizes are ideally independent of the profile shape, galaxy brightness, redshift, *global* luminosity evolution, and (in the absence of color gradients) the observed passband (Sandage & Perelmuter 1990). However, they are affected by noise, seeing, and for circular aperture photometry, galaxy ellipticity. Ideal axisymmetric $r^{1/4}$ profiles contain 87% of the total light within the metric radius; exponential disks include 95%.

Image shapes were measured using this definition for the Petrosian magnitude, $m_P$, with a shape parameter known as the image concentration index and defined (*e.g.*, by Kent 1985) as $C \equiv 5\log(r_{0.8}/r_{0.2})$, where the radius $r_N$ encloses a fraction $N$ of $m_P$. An axisymmetric $r^{1/4}$-law profile has $C = 4.6$ vs. 2.7 for an exponential disk; in comparison, $C = 2.1$ for a Gaussian and 1.5 for a uniform disk of finite radius. Although $C$ is a function of galaxy ellipticity, the dependence is small: an $r^{1/4}$ profile with axial ratio $b/a = 0.2$ has $C$ lower by about 0.2. Measurements of $C$ and $m_P$ for each galaxy appear in Table 1.

### 3.3. Simulations

Our aim is to separate early- and late-type galaxies (i.e., bulge- and disk-dominated systems) via the concentration index $C$ described above. Although under ideal circumstances pure disk and bulge systems would have the correct theoretical values of $C$, we do not know *a priori* the expected values for our deconvolved, noisy, and potentially unresolved galaxies. Simulations of galaxies provide an *a posteriori* path to this information, and so we measured $C$ for synthetic galaxies. Axisymmetric exponential disk or $r^{1/4}$-law profiles were convolved with the image of a deconvolved star to represent both spherical aberration and deconvolution, randomly inserted into our WFC image, and analyzed. No photon noise was added, since sky and readout noise dominated the random errors.

The simulated $C$ vs. $m_P$ relation appears in Figure 2(a). Whereas exponential disk profiles are close to the ideal value $C = 2.7$, pure bulge systems lie below the expected location $C = 4.6$ for a perfect $r^{1/4}$ profile, possibly due to imperfect image restoration that blurs the galaxy core, and pixellation which smooths the image. The latter effect is seen in the $C$ values for larger $r^{1/4}$ model galaxies being closer to 4.6. A dependence of $C$ on $m_P$ is especially evident for the $r^{1/4}$-law profiles. The observed trend towards lower $C$ at fainter $m_P$ is probably due to underestimation of the $r(\eta)$ radius; noise in the $\eta(r)$ profile causes the value $\eta = 0.1$ to be encountered at smaller radii. Bulge systems are more affected since they have more light in the outer part of the profile (their $\eta$ profiles are flatter at large $r$).

Based upon the simulations shown in Figure 2(a), we conclude that disk and bulge systems are well separated for $m_P \lesssim 21$, and can be discriminated past 22 on a statistical basis.

## 4. Discussion

In their study of distant clusters, DG divided the Cl0016+16 galaxies into three spectroscopic classes: late-type "k" spectrum galaxies, emission-line "e" spectrum objects, and an "a" class having a late-type "k" appearance plus strong Balmer absorption lines. Figure 2(b) displays our measurements for both cluster and field galaxies identified spectroscopically in the Cl0016+16 WFC field (DG).

The k-type galaxies have spectra resembling late-type K stars and are expected to be early-type E/S0 galaxies with old stellar populations. Figure 1 shows that "k" galaxies appearing in the top three rows are visually consistent with round, smooth morphology, except for the elongated galaxy #114. Few mergers or interactions appear in the group: 9 galaxies are normal or elongated, and only 4 have apparent companions. Quantitatively, the "k" galaxies lie primarily in the same region of the $C - m_P$ diagram as the simulated $r^{1/4}$ profiles (see Fig. 2(b)). The lower $C$ values ($C < 4$) of the 3 brightest "k" members result from extended halos and lower central surface brightness in these giant galaxies. The remaining two "k" galaxies with low $C$ (#205a, #439) are quite elongated. Some larger galaxies have higher $C$ than the simulated $r^{1/4}$ profiles, apparently because the real galaxies' cores are better resolved. Overall, the WFC images suggest the "k" objects are early-type bulge-dominated galaxies.

Galaxies typed "e" by DG show [O II], [O III], H$\alpha$, or H$\beta$ emission and hence are potential late-type spiral galaxies. The four "e" galaxies shown in the bottom row of Fig. 1 appear as low-surface-brightness objects with unusual morphology: as indicated in Table 1, each is elongated or double, and two have possible companions. Visually the "e" and "k" galaxies have different morphology, and our quantitative measurements show that they also have different shapes. For $m_P < 21$, the "e" and "k" types overlap little in the $C - m_P$ plane; i.e., both "e" objects (#186, #338) have lower $C$ values than any of the "k" objects except the very elongated galaxy (#114). The simulations suggest these two "e" galaxies are disk-dominated systems. The two "e" members with $m_P > 21$ (#413, #356) have $C$ values consistent with either bulge or disk morphology (although the high elongation for #356 implies it is not E/S0). Field galaxies (open symbols in Fig. 2(b)) support the correlation of "e" galaxies with disk morphology, since these objects (expected to consist primarily of spirals) have $C$ values consistent with the simulated disks.

Finally, the "a" galaxies have spectra combining a "k" appearance with strong Balmer absorption lines. The WFC images reveal a primarily disklike nature among the 8 objects of type "a" (with both Balmer and metal lines) or "A" (Balmer dominated). No visual characteristics distinguish the "a" and "A" galaxies, so we consider them together. Two have moderate-to-high surface brightness (#181, #352) and resemble normal bulges, but #352 may exhibit substructure and was classified as "uncertain a-type" in DG, and #181 is unfortunately bisected by a bad CCD column. The 6 lower-surface-brightness objects all have unusual shapes, but only two (#371, #411) show solid evidence for mergers or interactions. The other four "a" galaxies have irregular shape but no definite abnormalities. The region occupied by Cl0016+16's "a" galaxies in Fig. 2, like that for the "e" class, agrees with the model disks and is generally distinct from the "k." Whether these "e" and "a" galaxies are actually disk systems, or whether peculiarities of their profiles cause their lower $C$ values, these galaxies, on average, clearly differ in shape from the "k" cluster members, and are unlikely to be normal ellipticals.

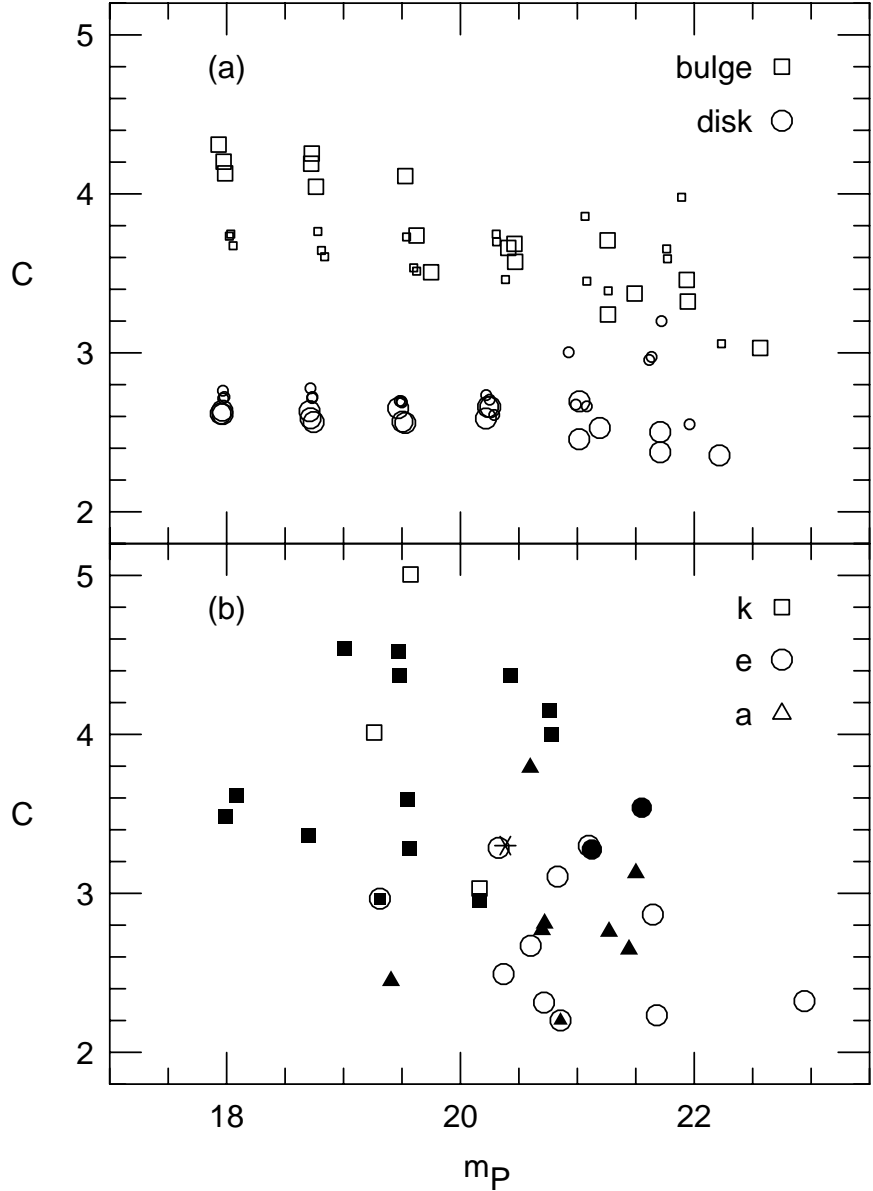

Fig. 2.— Concentration index $C \equiv 5\log(r_{80}/r_{20})$, where the $r_N$ radius encloses a fraction $N$ of our metric measure of the light (§3.2.), vs. *HST* Petrosian magnitude $m_P$. **(a)** Measurements for axisymmetric model galaxies convolved with the deconvolved PSF, inserted into the Cl0016+16 image, and analyzed. Larger symbols denote longer scale lengths for model galaxies. **(b)** Spectroscopically classified galaxies within the Cl0016+16 *HST* field, including both cluster members (filled symbols) and non-members (i.e., field; open symbols) with spectral types as indicated. Two member galaxies have composite spectra with emission lines, as indicated by the circled symbols. The asterisk represents object #181, an a-type cluster member contaminated by a bad CCD column and hence uncertain. Scatter here is higher than in the simulations due to greater size variation, non-axisymmetry, and the presence of galaxies which have both disk and bulge components.

These *HST* measurements of cluster galaxy morphology can be compared with visual findings from ground-based work (*e.g.*, Lavery, Pierce, & McClure 1992) and other *HST* programs (Couch *et al.* 1994, Dressler *et al.* 1994). Lavery *et al.* suggested interacting systems as the source of the excess blue galaxies reported in distant ($0.2 \lesssim z \lesssim 0.4$) galaxy clusters. Their recent observations of $z \approx 0.5$ clusters support this view; in particular, CFHT HRCAM images of Cl0016+16 show evidence for recent or ongoing interactions in most of the "e" and "a" galaxies (R. Lavery, private communication). Dressler *et al.* reach a similar conclusion by studying the $z = 0.41$ cluster 0939+4713. Based on assigning Hubble types to the cluster galaxies in both raw and deconvolved WFC-1 images, they found that all but one of 9 "e" or "a" members show evidence for recent mergers or galaxy encounters. Among the 18 "k" galaxies, only one showed definite morphological evidence for interactions. The Dressler *et al.* results thus support the Lavery *et al.* explanation.

These findings differ from those of Couch *et al.* using WFC-1 images of AC114 ($z = 0.31$). Comparison to the Couch *et al.* results is complicated by their different spectroscopic classification scheme. Whereas DG "a" galaxies have Hδ equivalent widths $W_\lambda(H\delta) > 4\text{Å}$, Couch *et al.* define two types with significant Hδ absorption based on color and equivalent width: a "post-starburst galaxy" (PSG) class comprising blue galaxies with $W_\lambda(H\delta) > 4\text{Å}$, and an "Hδ strong" (HDS) class of red galaxies with $W_\lambda(H\delta) > 2\text{Å}$. The Couch *et al.* "HDS" and "PSG" samples cover a similar range in $(g - r)$ color to the DG "a" class, suggesting "PSG" and "HDS" galaxies with $W_\lambda(H\delta) > 4\text{Å}$ could be considered "a" galaxies. Defined this way, the "a" galaxies of Couch *et al.* include three galaxies with bulge morphology, one disk, and an intermediate-morphology object. If all "PSG" and "HDS" objects are considered "a" galaxies, they include five bulge-dominated systems, 2 intermediates, and one disk. Depending on the preferred definition, either 3/5 or 5/8

"a" galaxies in AC114 feature bulge morphology, and only 1 or 2 show interactions or mergers. Hence, the isolated, bulge-dominated "a" galaxies in AC114 differ markedly from the disk-like, often interacting a-types seen in Cl0939+47 and Cl0016+16.

Although few ($\sim 20$) "E+A" objects have been surveyed thus far, the results show that the "a" galaxies in one cluster (AC114) differ considerably from those in other clusters (Cl0016+16 and Cl0939+47). We interpret this heterogeneity among the "E+A" populations in galaxy clusters as evidence that such galaxies may have diverse origins. The contrast between the bulge-dominated "E+A" galaxies in AC114 and the disk-like "E+A" objects in Cl0939+47 and Cl0016+16 hints that their morphology may vary systematically from cluster to cluster; however, the present sample of such galaxies is neither large enough nor sufficiently well defined to permit a firm statement regarding such dependence.

## 5. Conclusions

We have demonstrated that a quantitative analysis of *HST* imaging in distant clusters is practical and can reveal useful morphological information on galaxies. Specifically, the concentration index derived from the light growth curve provides a reliable first-order estimate of the profile shape, sufficient to distinguish between disk- and bulge-dominated galaxies in our WFC-1 images of Cl0016+16. Simulations show that galaxies with late-type K star ("k") spectra in this cluster have shapes typical of nearby elliptical galaxies, whereas emission-line ("e") and "E+A" ("a") galaxies are, on average, distinct and consistent with exponential disk profiles. Visual inspection of the images confirms these conclusions. Similar results were found in Cl0939+47, but not in AC114. Since "E+A" galaxies do not form a homogeneous morphological class, their origins may be diverse.

We are indebted to I. R. King for generously providing the data, to A. C. Phillips for advice on image restoration, and to M. A. Bershady for supplying software to measure galaxy profiles. The authors acknowledge grant support from CalSpace (CS 83-92), the UCSC Committee on Research, and NSF (AST-8858203).

---